\documentclass{elsart3p}
\usepackage{graphicx}
\usepackage{graphics}
\usepackage{amssymb,amsmath}

\begin{document}

\begin{frontmatter}

% Title, authors and addresses

% use the thanksref command within \title, \author or \address for footnotes;
% use the corauthref command within \author for corresponding author footnotes;
% use the ead command for the email address,
% and the form \ead[url] for the home page:
% \title{Title\thanksref{label1}}
% \thanks[label1]{}
% \author{Name\corauthref{cor1}\thanksref{label2}}
% \ead{email address}
% \ead[url]{home page}
% \thanks[label2]{}
% \corauth[cor1]{}
% \address{Address\thanksref{label3}}
% \thanks[label3]{}

\title{Tunneling Hamiltonian description of the atomic-scale 0-$\pi$ transition in superconductor/ferromagnetic-insulator junctions}

% use optional labels to link authors explicitly to addresses:
% \author[label1,label2]{}
% \address[label1]{}
% \address[label2]{}

\author[address1,address2]{S. Kawabata},
\author[address3]{Y. Tanaka},
\author[address4]{A. A.~Golubov},
\author[address5]{A. S.~Vasenko},
\author[address6]{S. Kashiwaya}
\author[address7]{Y. Asano}

\address[address1]{Nanosystem Research Institute (NRI), National Institute of Advanced Industrial Science and Technology (AIST), Tsukuba, Ibaraki 305-8568, Japan}

\address[address2]{CREST, Japan Science and Technology Corporation (JST), Kawaguchi, Saitama 332-0012, Japan}

\address[address3]{Department of Applied Physics, Nagoya University, Nagoya, 464-8603, Japan}

\address[address4]{Faculty of Science and Technology, University of Twente, P.O. Box 217, 7500 AE Enschede, The Netherlands}

\address[address5]{Institut Laue-Langevin, 6 rue Jules Horowitz, BP 156, 38042, Grenoble, France}

\address[address6]{Nanoelectronics Research Institute (NeRI), AIST, Tsukuba, Ibaraki 305-8568, Japan}

\address[address7]{Department of Applied Physics, Hokkaido University, Sapporo, 060-8628, Japan}

\begin{abstract}
We show a perturbation theory of the Josephson transport through ferromagnetic insulators (FIs).
Recently we have found that  the appearance of the atomic scale 0-$\pi$ transition in such junctions based on numerical calculations.
In order to explore the mechanism of this anomalous transition, we have analytically calculated the Josephson current using the tunneling Hamiltonian theory and found that the spin dependent $\pi$-phase shift in the FI barrier gives  the atomic scale 0-$\pi$ transition.
% We also show an experimental setup for observing the atomic-scale 0-$\pi$ transition based on a $d$-wave junction with a LBCO barrier.
\end{abstract}

\end{frontmatter}

% main text
\newpage
\section{Introduction}
In the usual Josephson junctions at equilibrium the phase difference of the superconducting order parameter on the two superconductor is zero. 
On the other hand, in the Josephson junctions with ferromagnetic-metal interlayer (S-MF-S junctions),  the ground state may correspond to the ƒÎ  phase difference~\cite{rf:Golubov,rf:Buzdin1}.
The Josephson $\pi$-junction formation  is related with the damping oscillatory behavior of the Cooper pair wave function in a FM~\cite{rf:Bulaevskii,rf:Buzdin2}. 
 In terms of the Josephson relationship $I_J= I_C \sin \phi$, where $\phi$ is the phase difference between the two superconductor layers, a transition from the 0 to $\pi$ states
implies a change in sign of $I_C$ from positive to negative. 
Experimentally the existence of the $\pi$-junction in S/FM/S systems has been confirmed by Ryanzanov et al.~\cite{rf:Ryanzanov} and Kontos et al.~\cite{rf:Kontos} 
Until recently, however,  investigations on the $\pi$ junction have been mainly focused on the S/FM/S systems.

We have predicted a possibility of the $\pi$-junction formation in Josephson junctions through {\it ferromagnetic insulators} (FIs) by numerically solving the Bogoliubov-de Genne equation~\cite{rf:Kawabata1,rf:Kawabata2} and the Nambu Green's function~\cite{rf:Kawabata3,rf:Kawabata4,rf:Kawabata5,rf:Kawabata6}.
The formation of the $\pi$ junction using such an insulating barrier is very promising for future qubit application~\cite{rf:Kawabata7,rf:Kawabata8,rf:Kawabata9,rf:Kawabata10} because of it's low decoherence nature~\cite{rf:Zaikin,rf:Kato}.
More importantly, we have shown that the ground state of such junction alternates between 0- and $\pi$-states when thickness of FI
is increasing by a single atomic layer.
In this paper in order to understand the physical mechanism of this atomic scale 0-$\pi$ transition, we analytically calculate the Josephson current based on the tunneling Hamiltonian method and show that the spin-dependent $\pi$-phase shift of the tunneling matrix element of the FI layer is the origin of this anomalous transition.

\section{Tunneling properties of a ferromagnetic insulator}
\begin{figure}[b]
\begin{center}
\includegraphics[width=8.0cm]{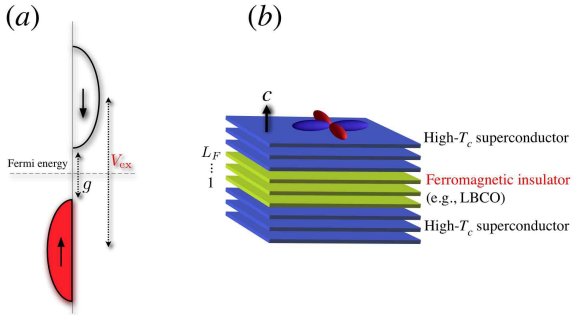}
\end{center}
\caption{(Color online) (a) The density of states for each spin direction for a ferromagnetic insulator, e.g., LBCO, and (b) schematic picture of $c$-axis stack high-$T_c$ $d$-wave superconductor/LBCO/high-$T_c$ $d$-wave superconductor Josephson junction. }
\label{fig1}
\end{figure}
In this section, we briefly describe the electronic structure of a representative of FI materials, i.e., La${}_2$BaCuO${}_5$ (LBCO)~\cite{rf:Mizuno}.
The typical DOS of LBCO for each spin direction is shown schematically in Fig. 1(a).
The exchange splitting $V_\mathrm{ex}$ is estimated  to be 0.34 eV by a first-principle band calculation using the spin-polarized local density approximation~\cite{rf:LBCO}.
Since the exchange splitting is large and the bands are originally half-filled, the system becomes FI. 
Based on the band structure [Fig. 1(a)], the spin $\sigma(= \uparrow, \downarrow)$ dependent transmission coefficient  $T_\sigma$ for the FI barrier can be calculated as~\cite{rf:Kawabata1}
\begin{eqnarray}
T_\sigma
=
\alpha_{L_F} \left( \frac{\rho_\sigma t}{g} \right)^{L_F}
,
 \end{eqnarray}
by using the transfer matrix method, where $\rho_{\uparrow(\downarrow)}=-(+)1$ and $\alpha_{L_F}$ is a spin-independent complex constant, $t$ is the hopping integral in the FI, $g$ is the gap between the up- and down-spin band, and $L_F$ is the layer number of the FI.
We immediately find 
\begin{eqnarray}
\frac{T_\uparrow }{T_\downarrow}
=
(-1)^{L_F},
 \end{eqnarray}
so the relative phase of $T_\sigma$ between spin up and down is $\pi$ (0) for the odd (even) number of $L_F$.
Next section, we will calculate the Josephson current through such a FI analytically.

In this paper we focused on the high-$T_C$ $d$-wave junction with a FI barrier [Fig. 1(b)].
We note that the qualitatively same result can be obtained for the case of conventional $s$-wave junctions.

\section{Theory}

In this section, we analytically calculate the Josephson current of S/FI/S junctions based on the tunneling Hamiltonian approach.
Let us consider a three-dimensional S/FI/S junction as shown in Fig.~1(b).
The Hamiltonian of the system can be described by 
\begin{eqnarray}
H = H_1 +H_2 +H_T +H_Q 
,
\end{eqnarray}
where $H_1$ and  $H_2$ are Hamiltonians describing the $d$-wave superconductors:
\begin{eqnarray}
H_{1}
&= &
\sum_\sigma \int d \mbox{\boldmath $r$} 
\ 
\psi_{1 \sigma}^\dagger \left( \mbox{\boldmath $r$}  \right)
\left( - \frac{\hbar^2 \nabla^2 }{2 m}  - \mu \right)
\psi_{1 \sigma}\left( \mbox{\boldmath $r$}  \right)
\nonumber\\
&-&
\frac{1}{2} \sum_{\sigma,\sigma'}  
\int d \mbox{\boldmath $r$}  \int  d \mbox{\boldmath $r$}'
\psi_{1 \sigma}^\dagger \left( \mbox{\boldmath $r$}  \right)
\psi_{1 \sigma'}^\dagger \left( \mbox{\boldmath $r$}'  \right)
\nonumber\\
&\times&
g_{1} \left( \mbox{\boldmath $r$}  - \mbox{\boldmath $r$}' \right)
\psi_{1 \sigma'} \left( \mbox{\boldmath $r$}'  \right)
 \psi_{1 \sigma} \left( \mbox{\boldmath $r$}  \right)
,
 \end{eqnarray}
where $\mu$ is the chemical potential and $\psi$ ($\psi^\dagger$) is the fermion field operator and $g \left( \mbox{\boldmath $r$}  - \mbox{\boldmath $r$}' \right)$ is the attractive interaction.
The tunneling Hamiltonian with a spin-dependent tunneling matrix element $t_\sigma$  of the FI barrier is given by 
\begin{eqnarray}
H_T
&= &
\sum_\sigma \int d \mbox{\boldmath $r$} \int  d \mbox{\boldmath $r$}'
\ 
\left[
t_\sigma \left( \mbox{\boldmath $r$},\mbox{\boldmath $r$}' \right) 
\psi_{1 \sigma}^\dagger \left( \mbox{\boldmath $r$}  \right)
\psi_{2 \sigma} \left( \mbox{\boldmath $r$}' \right)
\right.
\nonumber\\
&+&
\left.
\mbox{h.c.}
\right]
,
 \end{eqnarray}
and 
\begin{eqnarray}
H_Q
=
\frac{\left( Q_1 - Q_2 \right)^2 }{8C}
\end{eqnarray}
is the charging Hamiltonian, where $C$ is the capacitance of the junction and $Q_{1(2)}$ is the operator for the charge on the superconductor 1 (2), which can be written as 
\begin{eqnarray}
Q_{1(2)}
= 
e \sum_\sigma \int d \mbox{\boldmath $r$}
\psi_{1(2) \sigma}^\dagger \left( \mbox{\boldmath $r$}  \right)
\psi_{1(2) \sigma} \left( \mbox{\boldmath $r$}  \right).
\end{eqnarray}

By using the functional integral method~\cite{rf:Zaikin} and taking into account the spin dependence of $t_\sigma$ explicitly, the ground partition function for the system can be written as follows
\begin{eqnarray}
Z
=
\int  D \bar{\psi}_1D \psi_1 D \bar{\psi}_2 D \psi _2
\exp
\left[
  - \frac{1}{\hbar} \int_{0}^{\hbar \beta} d \tau 
    L (\tau) 
\right]
,
\nonumber\\
\end{eqnarray}
where $\beta=1/k_B T$, $ \psi(\bar{\psi})$ is the Grassmann field which corresponds to the fermionic field operator $[\psi (\psi^\dagger)]$ in Eq. (2), and the Lagrangian $L$ is given by
\begin{eqnarray}
	  L (\tau) 
	  =
	  \sum_\sigma \sum_{i=1,2} \int d \mbox{\boldmath $r$} 
	 \bar{\psi}_{i \sigma} \left( \mbox{\boldmath $r$}  , \tau \right) 
    \partial_\tau 
	\psi _{i \sigma} \left( \mbox{\boldmath $r$}  , \tau \right) 
	 + H (\tau) 
	 .
	 \nonumber\\
\end{eqnarray}
In order to remove the  $\psi^4$ term in the Hamiltonian $H( \tau)$, we will use the Hubbard-Stratonovich transformation which introduces a complex pair potential field 
\begin{eqnarray}
\Delta  ( \mbox{\boldmath $r$} , \mbox{\boldmath $r$}'  ;\tau)=\left| \Delta  ( \mbox{\boldmath $r$} , \mbox{\boldmath $r$}'  ;\tau)\right| \exp \left[ i \phi\left(   \mbox{\boldmath $r$} , \mbox{\boldmath $r$}'  ;\tau \right) \right].
\end{eqnarray}
The resulting action is only quadratic in the Grassmann field, so that the functional integral over this number can readily be performed explicitly.
The functional integral over the modulus of the pair potential field is taken by the saddle-point method.
Then the partition function is reduced to a single functional integral over the phase difference $\phi=\phi_1-\phi_2$.
To second order in the tunneling matrix element, one finds
\begin{eqnarray}
Z
= 
\int 
D \phi (\tau) 
\exp
\left[
  -  \frac{S_{\mathrm{eff}}[\phi] }{\hbar}
\right]
,
\end{eqnarray}
where the effective action is given by 
\begin{eqnarray}
S_{\mathrm{eff}}[\phi]
&=&
\int_{0}^{\hbar \beta} d \tau 
\left[
\frac{C}{2}
\left(
   \frac{\hbar}{2 e} \frac{\partial \phi ( \tau) }{\partial \tau}
\right)^2
   - \frac{\hbar}{2 e} I_C \cos \phi (\tau)
 \right]
.
\nonumber\\
\label{eqn:action}
\end{eqnarray}
In the calculation we have ignored the irrelevant quasiparticle tunneling term for simplicity.
Here 
\begin{eqnarray}
I_C
= -
\frac{2e }{\hbar} \int_{0}^{\hbar \beta} d \tau \beta(\tau)
\end{eqnarray}
is the Josephson critical current, and then the Josephson current is given by 
\begin{eqnarray}
I_J(\phi)= I_C \sin \phi
,
\end{eqnarray}
where the negative (positive) $I_C$ corresponds to the $\pi$ (0) junction.
The Josephson kernel $\beta(\tau)$ is given in terms of the  off-diagonal Nambu Green's function for two superconductor $F_i$ ($i=1,2$),
 $i. e.,$
\begin{eqnarray}
  \beta (\tau ) =
  -\frac{2}{\hbar}
  \sum_{\mbox{\boldmath $k$},\mbox{\boldmath $k$}'}
t_\downarrow^* (\boldsymbol{k},\boldsymbol{k'})
t_\uparrow (\boldsymbol{k},\boldsymbol{k'})  F_1 \left( \mbox{\boldmath $k$},\tau \right)
 F_2^\dagger \left( \mbox{\boldmath $k$}',-\tau \right)
 .
 \nonumber\\
\end{eqnarray}
The Nambu Green functions is given by 
\begin{eqnarray}
 F_i \left( \mbox{\boldmath $k$},\omega_n \right)
=
\frac{\hbar \Delta_i (\mbox{\boldmath $k$})}{
  (\hbar \omega_n)^2 + \xi_{\mbox{\boldmath $k$}}^2 +  \Delta_i (\mbox{\boldmath $k$})^2 }
,  
\end{eqnarray}
where $\xi_{\mbox{\boldmath $k$}}=\hbar^2 \mbox{\boldmath $k$}^2/2 m -\mu$ is the single particle energy relative to the Fermi surface and $\hbar \omega_n= (2n+1) \pi /\beta$ is the Matsubara frequency ($n$ is an integer).
In the case of the cuprate high-$T_c$ superconductors with the $d_{x^2-y^2}$ symmetry ~\cite{rf:Tanaka}, the order parameter is given by 
\begin{eqnarray}
\Delta_i (\mbox{\boldmath $k$})
  = 
  \Delta_0 \cos 2 \theta
  .
\end{eqnarray}
Below we will calculate the Josephson critical current $I_C$ analytically and discuss the possibility of the atomic scale 0-$\pi$ transition.

\section{Josephson critical current}

The Josephson critical current $I_C$ for a $c$-axis $d$-wave junction through the ferromagnetic insulator [Fig. 1(b)] can be expressed as 
\begin{eqnarray}
I_C
=
\frac{2 e^2  }{ \hbar \beta} 
\sum_{ \boldsymbol{k},\boldsymbol{k'}}
\sum_{\omega_n }
t_\downarrow^* (\boldsymbol{k},\boldsymbol{k'})
t_\uparrow (\boldsymbol{k},\boldsymbol{k'})
 F (\boldsymbol{k}, \omega_n) 
F (\boldsymbol{k'}, \omega_n)
\nonumber\\
  \end{eqnarray}
by assuming $\Delta_1=\Delta_2=\Delta_0 \cos 2 \theta$ and thus $ F_1 =  F_2=  F$.
We also assume that the tunneling matrix element $t_\sigma (\boldsymbol{k},\boldsymbol{k'}) \propto T_\sigma$ is given in terms of  the coherent tunneling in which the momentum $\boldsymbol{k}_\Vert$ parallel to the layer is conserved~\cite{rf:KawabataMQT1,rf:KawabataMQT4} and  has the same $L_F$ dependence on the transmission coefficient $T_\sigma$ as Eq. (1)~\cite{rf:Kawabata1}, $i.e$., 
\begin{eqnarray}
t_\downarrow^*  (\boldsymbol{k},\boldsymbol{k'}) 
t_\uparrow (\boldsymbol{k},\boldsymbol{k'})
 = |t_0|^2 (-1)^{L_F} \delta_{\boldsymbol{k}_\Vert \boldsymbol{k'}_\Vert }
, 
 \end{eqnarray}
we obtain an analytical expression of $I_C$ for $T=0$ K as 
\begin{eqnarray}
I_C =(-1 )^{L_F } \frac{\Delta_0 G_N }{2 \pi e}
,
 \end{eqnarray}
 where 
\begin{eqnarray}
G_N= \frac{4 \pi e^2 }{\hbar} |t_0|^2 N_0^2
,
 \end{eqnarray}
 is the normal conductance with $N_0$ being the density of states at $E_F$. 
 The sign of $I_C$ becomes negative for the odd number of $L_F$ and positive for the even number of $L_F$ as was numerically found in \cite{rf:Kawabata1,rf:Kawabata2,rf:Kawabata3,rf:Kawabata4,rf:Kawabata5,rf:Kawabata6}.
Therefore, the spin dependent $\pi$-phase shift of the tunneling matrix element $t_\sigma$ in the FI barrier gives rise to  the atomic scale 0-$\pi$ transition.

\section{Summary}
To summarize, we have studied the Josephson effect in S/FI/S junction by use of the tunneling Hamiltonian method.
We have analytically calculated the Josephson current and showed the possibility of the formation of the atomic scale 0-$\pi$ transition in such systems.
This observation is consistent with previous numerically results.
We hop that such FI based $\pi$-junctions become an  element in the architecture of quantum information devices.

\section*{Acknowledgements}

This work was  supported by CREST-JST, and a Grant-in-Aid for Scientific Research from the Ministry of Education, Science, Sports and Culture of Japan (Grant No. 22710096).

\end{document}